\documentclass[onecolumn,usenatbib,amsmath,amssymb,amsbsy]{mn2e}
\usepackage{amsmath,amssymb,mathrsfs,graphicx,float,latexsym,url}
\usepackage{epstopdf,ragged2e}
\usepackage{natbib}
\usepackage{hyperref}
\usepackage{varwidth}
\usepackage{longtable}
\usepackage{caption}
\usepackage{xtab}
\usepackage{multirow}
\usepackage[figuresright]{rotating}
\usepackage[usenames,dvipsnames]{color}
\usepackage{mathtools}
\usepackage{amsbsy}
\usepackage{bm}
\usepackage{float}
\usepackage{changes}
\usepackage{comment}
\usepackage{threeparttable}
\def\apj{{ApJ}}
\def\aj{{AJ}}

\def\aap{{A\&A}}

\def\mnras{{MNRAS}}

\def\nat{{Nature}}

\def\04a{{2004 a}}
\def\04b{{2004 b}}

\def\msun{{\rm M}_{\odot}}
\def\rsun{{\rm R}_{\odot}}
\def\mdot{\dot M}

\def\rco{R_{\rm {co}}}
\def\rlc{R_{\rm {lc}}}
\def\Rm{R_{\rm M}}
\def\ra{R_{\rm A}}
\def\ps{P_{\rm s}}
\begin{document}
	
	\title{Spin evolution of neutron stars in transient low-mass X-ray binaries
	}
	\author[Cui \& Li]{Zhe Cui$^{1,2,3}$
		and Xiang-Dong Li$^{1,2}$\thanks{Email: lixd@nju.edu.cn} \\
		$^1$School of Astronomy and Space Science, Nanjing University, Nanjing 210023, China\\
		$^2$Key Laboratory of Modern Astronomy and Astrophysics (Nanjing University), Ministry of Education, Nanjing 210023, China\\
		$^3$College of Physics and Electronic Information, Dezhou University, Dezhou 253023, China
	}
	
	\date{Accepted 2024 August 21. Received 2024 August 1; in original form 2024 March 22}
	
	\pagerange{\pageref{firstpage}--\pageref{lastpage}} \pubyear{?}
	
	\maketitle
	\label{firstpage}
	
	\begin{abstract}\label{sec:abs}
		
		\noindent Millisecond pulsar + helium white dwarf (MSP+He WD) binaries are thought to have descended from neutron star (NS) low-mass X-ray binaries (LMXBs). The NSs accreted from the progenitors of the WDs and their spin periods were accordingly accelerated to the equilibrium periods of order milliseconds. Thus, the initial spin periods of the ``recycled" NSs are critically determined by the mass transfer rate in the LMXB phase. However, the standard picture neglects the possible spin-down of the NSs when the donor star decouples from its Roche lobe at the end of the mass transfer, as well as the transient behavior of most LMXBs. Both imply more complicated spin evolution during the recycling process. In this work, we perform detailed calculations of the formation of MSP+He WD binaries. We take into account three magnetic braking (MB) prescriptions proposed in the literature, and examine the effects of both persistent and transient accretion. We find that the spin periods are not sensitively dependent on the efficiency of MB, but are considerably influenced by the accretion mode.  In comparison with persistent accretion, transient accretion leads to shorter and longer spin periods of the NSs in narrow and wide systems, respectively. This may help account for the measured spin periods of MSPs in wide binaries, which seem to be longer than predicted by the persistent accretion model.
		
	\end{abstract}
	
	
	%%%%%%%%%%%%%%%%%%%%%%%%%%%%%%%%%
	
	\section{Introduction}\label{sec:introduction}
	\label{sec:intro}
	Binary millisecond pulsars (MSPs) are neutron stars (NSs) characterised by rapid rotation (with spin periods $\ps\lesssim 30\, \rm {ms}$) and weak surface magnetic field strengths ($B_{\rm s}\lesssim 10^9\,\rm G$). They constitute a special subclass in the pulse period - period derivative diagram for radio pulsars, and differ markedly from ``normal" pulsars with relatively long pulse periods and strong magnetic field strengths \citep{Manchester2004Sci...304..542M}. In most cases, the companion stars of BMSPs are white dwarfs (WDs) with masses $\sim\,0.15-0.40\,\msun$. BMSPs are commonly thought to have descended from low-mass X-ray binaries (LMXBs), in which the NSs accrete mass and angular momentum from the companions that are overflowing from their Roche lobes (RLs).
	During the accretion phase, the NS magnetic fields are decayed and the spins are accelerated to milliseconds. This is the so-called recycling process \citep{Alpar1982Natur.300..728A, Radhakrishnan1982CSci...51.1096R, Bhattacharya1991PhR...203....1B}.
	After the cessation of mass transfer, the companions are left as He/CO WDs, and the recycled NSs turn to be radio MSPs. Then, the MSPs spin down and the WDs cool down more or less simultaneously \citep{Kulkarni1986ApJ...306L..85K}.
	
	The discovery of the first accreting millisecond X-ray pulsar (AMXP) SAX J1808.4$-$3658 (with the pulse period of 2.5 ms) by Rossi X-ray Timing Explorer  \citep{Wijnands1998Natur.394..344W} provides the first direct evidence of the recycling scenario. Until now the number of AMXPs has increased to 25 \citep{Papitto2022ASSL..465..157P, Sharma2023MNRAS.519.3811S}. The detection of the transitional behavior between accretion- and rotation-powered states in PSR J1023$+$0038, XSS J12270$-$4859, and IGR J18245$-$2452 \citep[e.g.,][]{Papitto2013Natur.501..517P, Bassa2014MNRAS.441.1825B, Archibald2009Sci...324.1411A} also demonstrates tight evolutionary link between MSPs and LMXBs.
	
	Theoretically, the MSP's spin-down is dominated by magnetic dipole radiation. Their ages can be calculated with \citep{Manchester1977puls.book.....M}
	\begin{equation}
		\tau_{\rm psr} =
		\frac{P_{\rm s}}{2\dot{P}_{\rm s}}\left[1-\left(\frac{P_{\rm s,i}}{P_{\rm s}}\right)^{2}\right],
		\label{eq:tpsr}
	\end{equation}
	where $P_{\rm s}$ is the spin period, $P_{\rm s, i}$ is its initial value, and $\dot{P}_{\rm s}$ is the period derivative. For $P_{\rm s,i}\ll P_{\rm s}$, the characteristic age $ \tau_{\rm c}\equiv P_{\rm s}/2\dot{P}_{\rm s}$ is used as an estimate of $\tau_{\rm psr}$.
	The spin-down age ($\tau_{\rm psr}$) of MSPs and the cooling age ($\tau_{\rm WD}$) of the WDs are independent and should approximately match each other. They can be used to test the MSP's rebirth period and probe the spin-down history \citep{Hansen1998MNRAS.294..569H}.
	
	Observations, however, reveal a more complex picture. For instance, \cite{van-Kerkwijk2000ApJ...530L..37V} predicted $\tau_{\rm WD}$ of PSR B1855+09's WD companion to be $\sim 10\,\rm Gyr$, which is twice as long as its $\tau_{\rm c}$. As to the spin period distribution, it was found that binary MSPs seem to spin slower than AMXPs on average \citep{Ferrario2007, Hessels2008AIPC.1068..130H,Papitto2014}. In principle, the NSs in LMXBs can be spun up to submillisecond periods, but both observed MSPs and AMXPs show a cutoff spin period around 1.3 ms \citep{Chakrabarty2003,Hessels2006Sci...311.1901H, Patruno2010}. Angular momentum loss via gravitational wave emission may be effectively spinning down the accreting NSs caused by quadrupole asymmetry in the star's moment of inertia \citep{Bildsten1998}
	or by the r-mode instability \citep{Andersson1998}.
	
	Alternatively, accretion torques should be also playing an important
	role to regulate the spin of accreting NSs \citep[see][for discussion]{Patruno2017,Patruno2021,Tauris2023}.
	It was noted that, at the end of Roche lobe overflow (RLOF), the companion gradually decouples from its RL, and with decreasing mass transfer rate, the accreting NSs are likely subject to considerable spin-down, deviating from their previous equilibrium spin periods \citep{Tauris2012Sci...335..561T}. Furthermore, observationally most LMXBs are found to be transient rather persistent X-ray sources, experiencing short outbursts separated by long quiescent intervals. The inferred duty cycle is on average $\sim 3\%$, ranging from below $1\%$ to above $10\%$ \citep{Yan2015ApJ...805...87Y}.
	The X-ray luminosity varies between outburst and quiescence by orders of magnitude, suggesting a big change in the accretion rate.
	With semi-analytic analysis, \citet{Bhattacharyya2017ApJ...835....4B} argued that NSs with transient accretion could spin up to shorter periods than derived from the average mass accretion rate.
	
	In this work, we combine detailed binary evolution with the accretion torque to investigate its influence on the NS spin evolution, and evaluate the initial periods of binary MSPs. We collect MSP+He WD binaries in which $\tau_{\rm WD}$ have been measured photometrically, and compare them with $\tau_{\rm psr}$ to test the models. The following of the paper is organized as follows. We introduce our numerical methods and the physical assumptions on the evolution of LMXBs in Section \ref{sec:methods}. Our calculated results of two representative examples are presented in Section \ref{sec:results}. We compare the results with observation and discuss their implications in Section \ref{sec:compare}, and finally summarize in Section \ref{sec:conclusion}.
	
	\section{Methods}\label{sec:methods}
	\subsection{Evolution of LMXBs}\label{subsec:evolution}
	
	We follow the binary evolution using the one-dimensional stellar evolution code Modules for Experiments in Stellar Astrophysics (MESA) (version 11701) \citep{Paxton2011ApJS..192....3P, Paxton2013ApJS..208....4P, Paxton2015ApJS..220...15P, Paxton2018ApJS..234...34P, Paxton2019ApJS..243...10P}. Each evolution begins from an incipient LMXB consisting of a NS and a zero-age main-sequence (MS) donor star, and terminates until the formation of a detached NS+He WD system within a Hubble time.
	For the initial binary parameters, we set the NS mass $M_{\rm NS}=1.3\,\msun$ and consider two different initial donor masses of $M_{2}\,=\,1.0$ and $1.5\,\msun$. The orbital periods $P_{\rm orb}$ at the beginning of the evolution range from the minimum value with which the systems would evolve to be NS+He WD binaries within a Hubble time, to the maximum value with which the final orbital period of the NS+He WD binary approaches $\sim 500\,\rm day$ (i.e., the maximum observed period of our selected MSP+He WD sample). We adopt solar metallicity for the donor star.
	The boundary of the stellar convective region is determined using the Schwarzschild criterion, and treated with the \cite{Bohm1958ZA.....46..108B} mixing-length theory with a mixing-length parameter of 1.82. For overshoot mixing we follow the treatment of \cite{Choi2016ApJ...823..102C}, where the overshooting efficiency is determined by two sets of parameters ($f_{\rm ov}$, $f_{\rm 0,ov}$) which are set to be (0.016, 0.008) for the core and (0.0174, 0.0087) for the envelope/shell during any kind of the burning phase.
	
	Mass transfer in LMXBs is usually nonconservative. We assume a $30\%$ accretion efficiency \citep{Tauris2012MNRAS.425.1601T,Antoniadis2012MNRAS.423.3316A}, that is, $\dot{M}_{\rm NS}\,=0.3\,\dot{M}_{\rm T}$, where $\dot{M}_{\rm T}$ and $\dot{M}_{\rm NS}$ are the mass transfer rate and the NS's accretion rate, respectively. The excess material is assumed to be lost from the NSs in the form of istropic winds. For supper-Eddington mass transfer, accretion is limited by the Eddington accretion rate ($\dot{M}_{\rm Edd}\simeq 2\times 10^{18}$ gs$^{-1}$ for a $1.3\,\msun$ NS).
	
	We consider loss of orbital angular momentum ($\dot{J}_{\rm orb}$) caused by gravitational wave radiation, mass loss, and magnetic braking as described in \cite{Paxton2015ApJS..220...15P},
	\begin{equation}
		\dot{J}_{\rm orb} = \dot{J}_{\rm gr} + \dot{J}_{\rm ml} + \dot{J}_{\rm mb}.
	\end{equation}
	
	Gravitational wave radiation dominates angular momentum loss for very short orbital period systems, which is well understood and observationally confirmed \citep{Kramer2006Sci...314...97K}. The loss rate is given by
	\begin{equation}
		\dot{J}_{\rm gr} = -\frac{32}{5c^5}\left(\frac{2\pi G}{P_{\rm orb}}\right)^{7/3}\frac{(M_{\rm NS} M_2)^2}{(M_{\rm NS}+M_2)^{2/3}},
	\end{equation}
	where $G$ and $c$ are the gravitational constant and the speed of light in vacuum, respectively.
	
	Mass loss from the donor and from the NS takes the specific orbital angular momenta of the donor and the NS, respectively. Thus, the corresponding angular momentum loss rate is
		\begin{equation}
			\dot{J}_{\rm ml} = -[\dot{M}_{2,\rm w}M_{\rm NS}^2 + (\dot{M}_{\rm T}-\dot{M}_{\rm NS})M_2^2]\frac{a^2}{(M_{\rm NS}+M_2)^2}\left(\frac{2\pi}{P_{\rm orb}}\right),
	\end{equation}
	where $a$ is the binary separation, and the stellar wind mass loss rate $\dot{M}_{2,\rm w}$ is calculated using the `Dutch wind' prescription with the scaling factor of 1.0.
	
	Magnetic braking (MB) dominates angular momentum loss in relatively compact LMXBs. However, the physical mechanism still remains unclear. Here we adopt three representative MB prescriptions in the literature to investigate its influence on the formation of MSP+He WD binaries.
	\begin{itemize}
		\item MB1: The commonly used \cite{Skumanich1972ApJ...171..565S} empirical MB prescription, with the parametrized form given by \citep{Rappaport1983ApJ...275..713R},
		\begin{equation}
			\dot{J}_{\rm mb,MB1}= -3.8\times 10^{-30}M_2\rsun^4(\frac{R_2}{\rsun})^{\gamma_{\rm mb}}\Omega_2^3,
		\end{equation}
		where $R_2$ and $\Omega_2$ are the radius and the angular velocity of the donor star respectively, and $\Omega_{\odot} = 3.0\times10^6\,\rm {s^{-1}}$. The index $\gamma_{\rm mb}$ is taken to be 4.
		\item MB2: The enhanced convective turnover time-scale MB prescription suggested by \citet{Van2019MNRAS.483.5595V},
		\begin{equation}
			\dot{J}_{\rm mb,MB2} = \dot{J}_{\rm mb,MB_1}(\frac{\tau_{\rm conv}}{\tau_{\odot,\rm conv}})^2,
		\end{equation}
		where $\tau_{\rm conv}$ and $\tau_{\odot,\rm conv}$ are the convection turnover times of the donor and the Sun, respectively.
		\item MB3: The convection and rotation boosted prescription in \citet{Van2019ApJ...886L..31V},
			\begin{equation}
				\dot{J}_{\rm mb, MB3} = -\frac{2}{3}\Omega_{\odot}B_{\odot}^{8/3}\dot{M}_{2,\rm w}^{-1/3}R_2^{14/3}(v^2_{\rm esc}+2\Omega_2^2R_2^2/K_2^2)^{-2/3} (\frac{\Omega}{\Omega_{\odot}})^{11/3}(\frac{\tau_{\rm conv}}{\tau_{\odot,\rm conv}})^{8/3},
		\end{equation}
		where $v_{\rm esc}$ is the surface escape velocity of the donor star, $K_2=0.07$ is a constant obtained from a grid of simulations by \cite{Reville2015ApJ...798..116R}, and $B_{\odot}\, = \,1$ G is the Sun's surface magnetic field strength.
	\end{itemize}

	\subsection{Spin evolution of the NSs}\label{subsec:spin}
	The inner part of the accretion disc around a magnetized NS can be truncated by the stellar magnetic field. The interaction between the NS's magnetic field and the disc and the corresponding torque exerted on the NS  depend on the relative size of the following three radii.
	\begin{itemize}
		\item The magnetospheric radius $\Rm =\phi\ra$, defined as the location where the magnetic field becomes strong enough to channel the incoming material to corotates with the NS. It also represents the inner edge of the disc. Here $\phi\,\simeq\,0.5-1.4$ is a dimensionless constant \citep{Ghosh1992ASIC..377..487G, Arons1993ApJ...408..160A} (we adopt $\phi=1.0$), and $\ra$ is the Alf\'{v}en radius
		\begin{equation}
			\ra = \left(\frac{\mu ^2}{\sqrt{2GM_{\rm NS}}\dot{M}_{\rm NS}}\right)^{2/7}
			\simeq
			3.2\times10^8({\rm cm})\mu_{30}^{4/7}m^{-1/7}\dot{M}_{17}^{-2/7},
			\label{eq:rm}
		\end{equation}
		where $\mu_{30}\,=\,\mu/(10^{30}\,{\rm Gcm^3})=B_{\rm s}R_{\rm NS}^3/(10^{30}\,\rm Gcm^3)$ is the dimensionless magnetic moment, $m=M_{\rm NS}/\msun$, and $\mdot_{17}=\dot{M}_{\rm NS}/10^{17}\,\rm gs^{-1}$.
		\item
		The corotation radius $\rco$, at which the Keplerian angular velocity in the disc is the same as the NS's angular velocity,
		\begin{equation}
			\rco = \left(\frac{GM_{\rm NS}}{\Omega_{\rm s}^2}\right)^{1/3}\simeq 1.5\times10^8({\rm {cm}}) m^{1/3}\ps^{2/3},
			\label{eq:rco}
		\end{equation}
		where $\Omega_{\rm s}=\ps/2\pi$ is is the angular velocity of the NS.
		\item The light cylinder radius $\rlc$,  given by
		\begin{equation}
			\rlc = \frac{c}{\Omega_{\rm s}}	\simeq 4.8\times10^9({\rm {cm}})\ps.
		\end{equation}
	\end{itemize}
	
	The NS transits among the following states along with the evolution of the mass transfer rate, the NS's magnetic field and spin period.
	\begin{itemize}
		\item [(1)] The accretor state\\
		If the disc is truncated inside $\rco$ and $\rlc$, that is, $\Rm<\rco<\rlc$, the NS can steadily accrete, and there is an external torque acting on the accreting NS consisting of two components,
		\begin{equation}
			N = N_{\rm acc} + N^+_{\rm {mag}},
		\end{equation}
		where
		\begin{equation}
			N_{\rm acc} = \dot{M}_{\rm NS}\sqrt{GM_{\rm NS}\Rm},
		\end{equation}
		is torque due to mass accretion, and
		\begin{equation}
			N^+_{\rm {mag}} =\frac{\mu^2}{9 \Rm^3}\left[2\left(\frac{\Rm}{\rco}\right)^3-6\left(\frac{\Rm}{\rco}\right)^{3/2}+3\right],
		\end{equation}
		is the torque caused by twisted magnetic field lines that thread the accretion disc.
		\item[(2)] The propeller state\\
		If the disc is truncated outside $\rco$ but within $\rlc$, that is, $\rlc>\Rm>\rco$, then most or all of the accreted material is expelled from the NS due to the centrifugal barrier. There is a negative torque on the NS caused by mass ejection and NS-disc interaction,
		\begin{equation}
			N = -\eta N_{\rm acc} + N^-_{\rm {mag}},
		\end{equation}
		where
		\begin{equation}
			N^-_{\rm {mag}} = -\frac{\mu^2}{9 \Rm^3}\left[3-2\left(\frac{\rco}{\Rm}\right)^{3/2}\right],
		\end{equation}
		and $\eta$ is a constant of order of unity, which describes the uncertainty due to  mass ejection. Here we adopt $\eta = 1$.
		\item[(3)] The radio pulsar state\\
		If $\Rm\geq\rlc$, the NS acts as a radio pulsar and spins down due to magnetic dipole radiation. The spin-down torque can be expressed as
		\begin{equation}
			N=N_{\rm EM} = -\frac{2\mu^2}{3 \rlc^3}.
		\end{equation}
	\end{itemize}
	There are various derivations of the magnetic torques $N^+_{\rm mag}$ and $N^-_{\rm mag}$ in the literature. Here we use the same expressions in \cite{Bhattacharyya2017ApJ...835....4B}.
	
	The spin period evolution of the NS follows the equation
	\begin{equation}
		I\dot{\Omega}_{\rm s} = N,
	\end{equation}
	where $I$ is the moment of inertia of the NS. When there is no net torque exerted onto an accreting NS, it reaches an equilibrium spin period,
	\begin{equation}
		P_{\rm s, eq} = \phi^{2/3}\mu^{6/7}(G^5M_{\rm NS}^5\dot{M}_{\rm NS}^3)^{-1/7}\simeq 1.4\, ({\rm ms})\,B_{8}^{6/7}(\frac{\dot{M}_{\rm NS}}{0.1\dot{M}_{\rm Edd}})^{-3/7}(\frac{M_{\rm NS}}{1.4\msun})^{-5/7}R_{6}^{18/7},
		\label{eq:peq}
	\end{equation}
	where $B_8=B_{\rm s}/10^8\,\rm G$ and $R_{6}=R_{\rm NS}/10^{6}\,\rm {cm}$.
	
	We take the common wisdom that the NS magnetic field rapidly decays due to mass accretion, and saturates around $\sim 10^8$ G. This is consistent with the fact in most MSPs $B_{\rm s}$ ranges from $3\times10^7$ to $4\times10^8\,\rm G$ as seen in Table \protect\ref{tab:tab1}. So we adopt three constant $B_{\rm s}$ values in our calculations, instead of considering the magnetic field evolution during the LMXB phase.
	
	\subsection{Transient accretion}\label{subsec:transient}
	All AMXPs and most NS LMXBs are transients \citep{Papitto2022ASSL..465..157P}. The most plausible interpretation is thought to be the occurrence of thermal-viscous instability in the accretion disc if the mass transfer rate is below a critical value $\dot{M}_{\rm cr}$ \citep{Dubus1999MNRAS.303..139D}:
	\begin{equation}
		\dot{M}_{\rm cr} = 3.2\times10^{-9}\left(\frac{M_{\rm NS}}{1.4\msun}\right)^{0.5}\left(\frac{M_{\rm 2}}{\msun}\right)^{-0.2}\left(\frac{P_{\rm orb}}{1 \rm day}\right)^{1.4} \msun \rm yr^{-1}\label{mdi}.
	\end{equation}
	An unstable disc would experience cycles of outbursts separated by long duration quiescence. The transferred material is accumulated in the disc during quiescence, and rapidly accretes during outbursts.
	We define $f$ to be the outburst duty cycle, that is, $f = t_{\rm out}/( t_{\rm out}+t_{\rm qui})$, in which $t_{\rm out}$ and $t_{\rm qui}$ are the durations of outbursts and quiescence, respectively. Observations suggest that $f$ ranges from $\sim 0.01-0.2$ for NS LMXBs \citep{King2003MNRAS.341L..35K, Yan2015ApJ...805...87Y}, so we take $f=0.01$ and $0.1$ in our calculations. According to \cite{Paradijs1996ApJ...464L.139V}, the quiescent duration in an outburst cycle can be estimated by
	\begin{equation}
		t_{\rm qui} = 0.5(\frac{\alpha_{\rm cold}}{0.01})^{-1}(\frac{\dot{M_{\rm T}}}{10^{17}~\rm gs^{-1}})^{-2}(\frac{T_c}{3000~\rm K})^{-1} (\frac{M}{\msun})^{-1.26}(\frac{R_{\rm disc}}{10^{10}~\rm cm})^{5.8}~\rm {day},
	\end{equation}
	where $ R_{\rm disc}$ is the outer disc radius which is about $70\%$ of the NS's RL radius, and $T_{\rm c}$ the disc temperature after considering X-ray irradiation (roughly equals to the hydrogen ionization temperature $\sim$ 6500 $\rm K$). Observations indicate that the recurrence times of LMXB transients range from several days to decades \citep{Lasota2001NewAR..45..449L, Maccarone2022MNRAS.512.2365M}. Note that all LMXBs are subject to the thermal-viscous instability when the donor is decoupling from its RL to end the mass transfer.
	
	The accretion rate of the NS is then set to be
		\begin{equation}
			\dot{M}_{\rm NS} = \left\{
			\begin{array}{ll}
				0.3\min(\dot{M}_{\rm T},\dot{M}_{\rm Edd}), & {\rm if}\ \dot{M}_{\rm T}\geq \dot{M}_{\rm cr}\\
				0.3\min(\dot{M}_{\rm T}/f,\dot{M}_{\rm Edd}),& {\rm if}\ \dot{M}_{\rm T}< \dot{M}_{\rm cr}, \rm \;{\; during\; outbursts}\\
				0 & {\rm if}\ \dot{M}_{\rm T}<\dot{M}_{\rm cr}, \;\rm {\; during\; quiescence}.
			\end{array}
			\right.\label{mdotns}
	\end{equation}

	\section{Results}\label{sec:results}
	\subsection{Initial parameters of the model grids}\label{subsec:initial paremeters}
	We create a grid of models for LMXBs consisting a NS of mass $M_{\rm NS}=1.3\,\msun$ and a donor star of initial mass $M_2=1.0$ and $1.5\,\msun$. We adopt three different MB laws (MB1, MB2 and MB3). The NS magnetic field strengths are taken to be $3\times10^7, \,2\times10^8\,\rm{and}\, 4\times10^8\, \rm G$, and the initial spin period to be $10\,\rm s$.  We explore both the `DI' and `stable' models with and without thermal-viscous disk instability considered, respectively. For each model, we try a range of initial orbital periods, yielding a roughly uniform distribution of $P_{\rm orb}$ for MSP+He WD binaries among $0.1-500\,\rm{day}$.
	
	\subsection{Examples of LMXB evolution}\label{subsec:examples}
\begin{figure*}
		\includegraphics[width=1.0\linewidth]{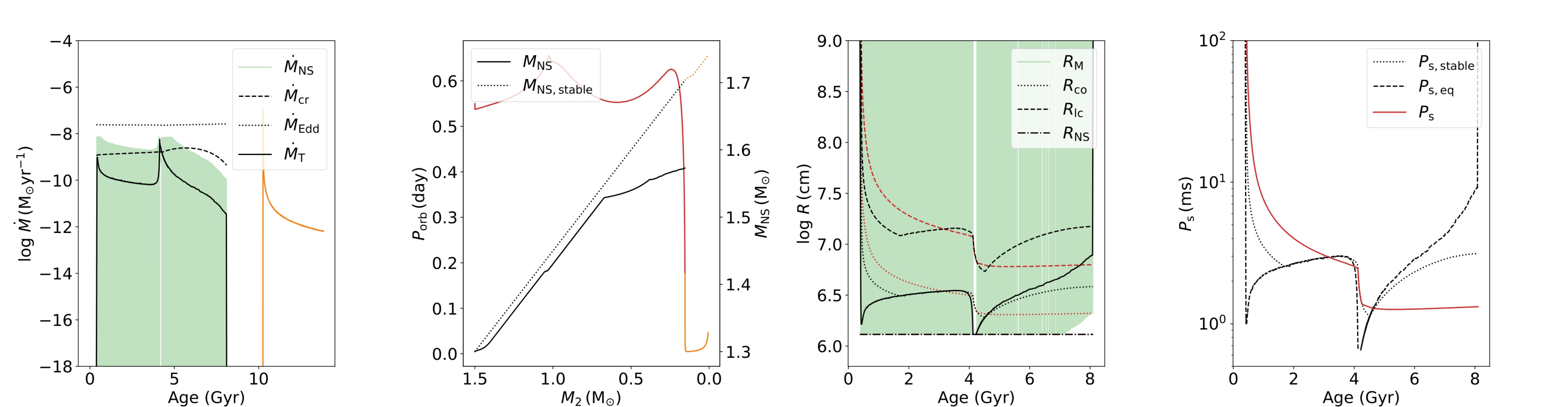}
		\includegraphics[width=1.0\linewidth]{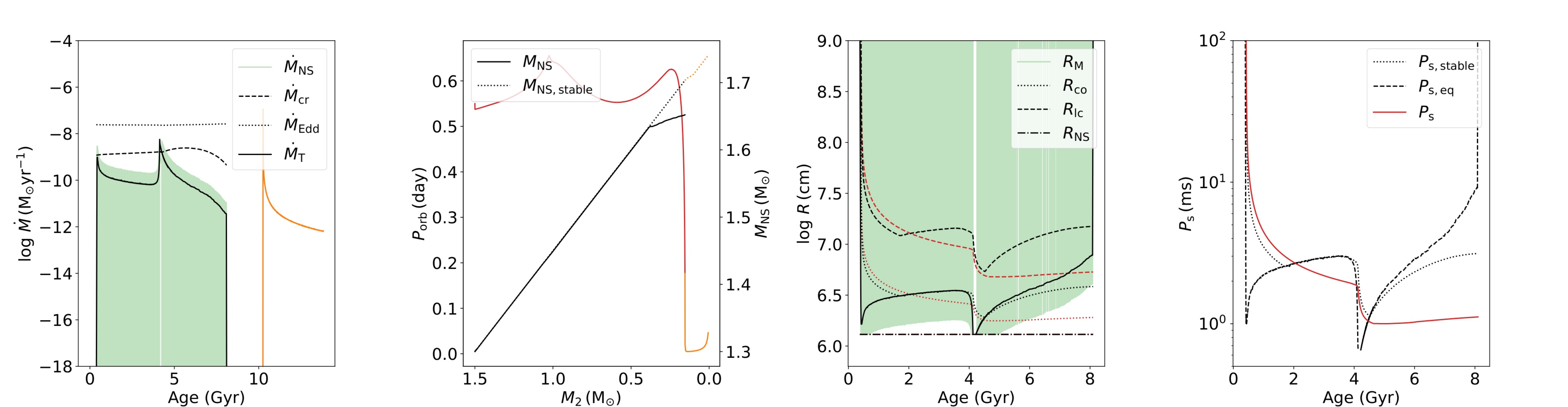}
		\caption{Evolution of a LMXB with the initial parameters ($M_2$, $M_{\rm NS}$, $P_{\rm orb}$) = ($1.5\,\msun$, $1.3\,\msun$, $0.55\,\rm day$). We adopt MB1 prescription and $B_{\rm s} = 2\times10^8\,\rm G$. The outburst duty cycle $f=0.01$ (top row) and $0.1$ (bottom row), respectively. In the first column, the black dotted, dashed and solid lines represent the Eddington mass-accretion rate $\dot{M}_{\rm Edd}$, critical mass transfer rate for disc instability $\dot{M}_{\rm cr}$ and mass transfer rate $\dot{M}_{\rm T}$ as the function of age, respectively. The shaded green region depicts the accretion rate changing alternatively between outbursts and quiescence in `DI' model. In the second column, the black dotted and solid lines represent the NS mass as a function of donor star mass in the `stable' and `DI' model, respectively. The red solid line represents orbital period. After the first RLO phase, the donor star settles as a He WD and fills its Roche lobe again, we show the evolution during this UCXB phase with yellow lines in the first two columns. In the third column, the black (red) dotted and dashed lines represent the variation of corotation radius and light cylinder radius of NS in the `stable' and `DI' model, respectively, and the radius of NS is shown with dash-dotted line in the bottom. The black solid line represents the magnetospheric radius of NS in the `stable' model, and in `DI' model is represented by green shaded region. The last column shows the evolution of the NS's spin period $P_{\rm s, stable}$ (black dotted line) and the equilibrium spin period $P_{\rm s,eq}$ (black dashed line) in the `stable' model, and spin period $P_{\rm s}$ (red solid line) in the `DI' model.} \label{fig:fig1}
	\end{figure*}

	In Fig.~\ref{fig:fig1} we present the evolution of a narrow LMXB with the initial parameters ($M_2,\, M_{\rm NS},\,P_{\rm orb}$) = ($1.5\,\msun,\,1.3\,\msun,\,0.55\,\rm day$), $B_{\rm s} = 2\times10^8\, \rm {G}$, and the MB1 prescription. The upper and lower panels depict the results with the duty cycle $f=0.01$ and 0.1, respectively.
	
	The first column of Fig.~\ref{fig:fig1} shows the evolution of the mass transfer rate $\dot{M}_{\rm T}$ (black solid line), the critical mass transfer rate $\dot{M}_{\rm cr}$ (black dashed line), the NS accretion rate $\dot{M}_{\rm NS}$ (green solid line), and the Eddington accretion rate $\dot{M}_{\rm Edd}$ (black dotted line) with the age. At the age of $\sim 4.0\,\rm Gyr$, the core of the donor star becomes radiative and thus MB begins to work, which causes a rapid increase in $\dot{M}_{\rm T}$. During most of the mass transfer process, $\dot{M}_{\rm T}$ is lower than $\dot{M}_{\rm cr}$, and the binary appears as a recurrent X-ray transient. The accretion rate changes alternatively between outbursts and quiescence, so we use the shaded green region to depict this stage, and only plot the accretion rate $\dot{M}_{\rm NS}$ during outbursts. We see that it is substantially larger than $\dot{M}_{\rm T}$, and can reach $\dot{M}_{\rm Edd}$ in the case of $f=0.01$. The He WD is born at the age of $8.1\,\rm Gyr$. After that, the system becomes detached and then contracts due to gravitation radiation, evolving to an ultra-compact X-ray binary (UCXB). We label the UCXB phase in the first and second columns with the yellow lines.
	
	The second column shows the evolution of the orbital period (red solid line) and the NS mass (black dotted and solid lines in the `stable' and `DI' models, respectively) with the donor star mass. Owing to transient accretion during outbursts, a less massive NS is produced in the `DI' model (the final mass $M_{\rm NS} = 1.575/1.65\,\msun$ with $f=0.01/0.1$).
	
	The third column compares the evolution of the magnetospheric radius (black and green solid lines in the `stable' and `DI' models respectively), the corotation radius (dotted line), the light cylinder radius (dashed line), and the NS's radius (dot-dashed line). For the corotation radius and the light cylinder radius, the black and red colors correspond to the `stable' and `DI' models, respectively.
	The last column demonstrates the evolution of the NS's spin period $P_{\rm s, stable}$ (black dotted line) and the equilibrium spin period $P_{\rm s,eq}$ (black dashed line) in the `stable' model, and $P_{\rm s}$ (red solid line) in the `DI' model.
	In the `stable' model, $\Rm<\rco$ soon after the mass transfer starts. The two radii coincide at the age of $1.6\,\rm Gyr$, and the NS accordingly reaches the equilibrium period ($P_{\rm s, stable}=P_{\rm s,eq}$). At the age of $4.0\,\rm Gyr$, the rapid increase of $\dot{M}_{\rm NS}$ due to the initiation of MB pushes $\Rm$ inside $\rco$, and causes the NS to spin up to near 1 ms\footnote{The spin period may be under-estimated here since we have neglected possible spin-down via gravitational radiation.}. After that, the donor star gradually decouples from its RL. With decreasing $\dot{M}_{\rm NS}$, the spin equilibrium is eventually broken because $\Rm$ expands more quickly than $\rco$. The propeller torque spins the NS from $1.0\,\rm ms$ down to $3.0\,\rm ms$. In the `DI' model, however, $\Rm$ (during outbursts) is almost always smaller than $\rco$ during outbursts. Although the NS is spun down during quiescence, the rapid accretion during outbursts is able to spin up the NS to a period of nearly $1.0\,\rm ms$. The slow decoupling phase has less effect on $P_{\rm s}$ when $f=0.01$, but when $f=0.1$, it also results a slightly spin-down of the NS.
	
\begin{figure*}
		\includegraphics[width=1.0\linewidth]{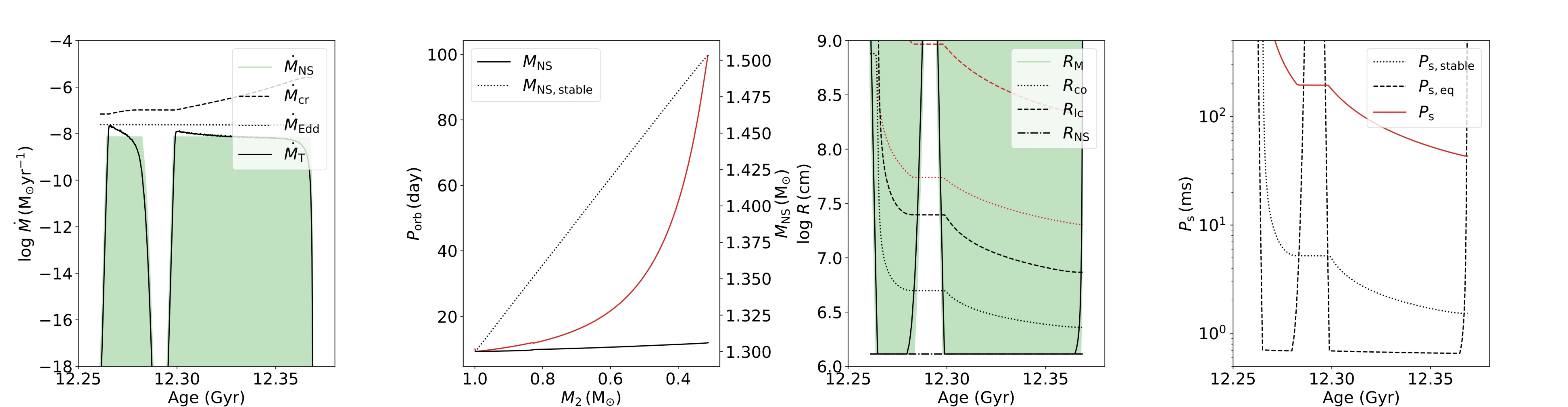}
		\includegraphics[width=1.0\linewidth]{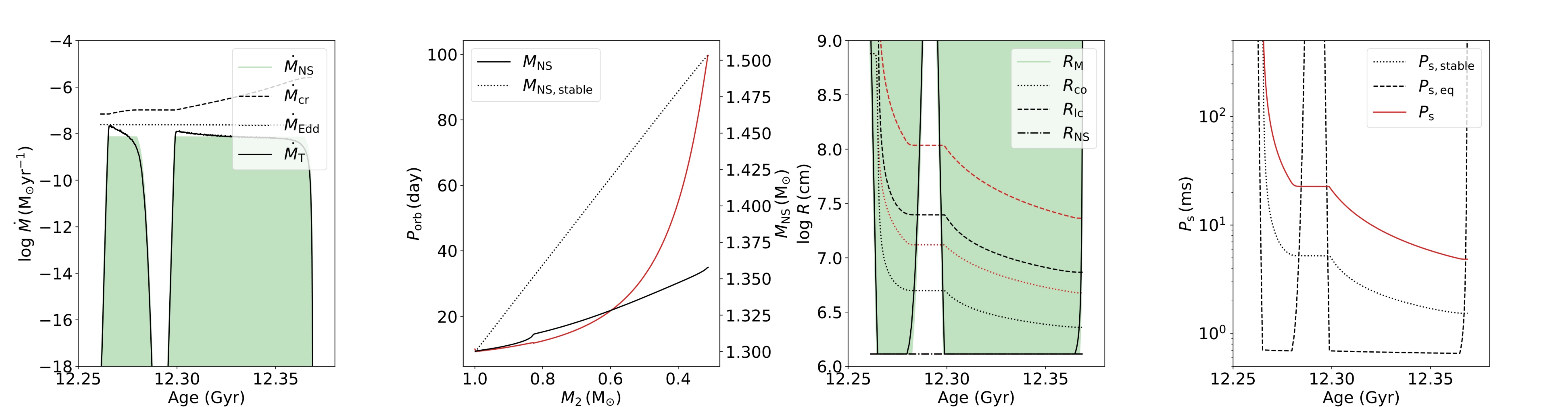}
		\caption{Same as Fig. \ref{fig:fig1}, but for initial parameters ($M_2$, $M_{\rm NS}$, $P_{\rm orb}$) = ($1.0\,\msun$, $1.3\,\msun$, $10.0\,\rm day$).}\label{fig:fig2}
	\end{figure*}

	Fig.~\ref{fig:fig2} shows the evolution of a wide LMXB with the initial parameters ($M_2$, $M_{\rm NS}$, $P_{\rm orb}$) = ($1.0\,\msun$, $1.3\,\msun$, $10.0\,\rm day$). As the donor star is more evolved in wide system before RLOF occurs, mass transfer proceeds on shorter timescale\footnote{There is a temporary detachment as shown in the $\dot{M}-\rm age$ diagram. It occurs when the H-burning shell crosses the discontinuity caused by the first dredge-up.}. The orbital period increases all the way to $\sim 100$ day. The accretion efficiency and spin evolution differ significantly between the `stable' and `DI' models. The accretion rate during outbursts is always super-Eddington, so the amount of mass growth is substantially smaller in the `DI' model than in the `stable'  model. In both models the magnetospheric radius $\Rm$ is smaller than $\rco$, the NS keeps spinning up to $P_{\rm s, eq}$, but the small accretion torque prohibits it reaching the equilibrium spin. The final spin periods are 1.5 ms in the `stable' model and 40.0/5.0 ms (with $f=0.01/0.1$) in the `DI' model.
	
	\section{Comparison with observations and discussion}\label{sec:compare}
	
\begin{figure*}
		\includegraphics[width=1.0\linewidth]{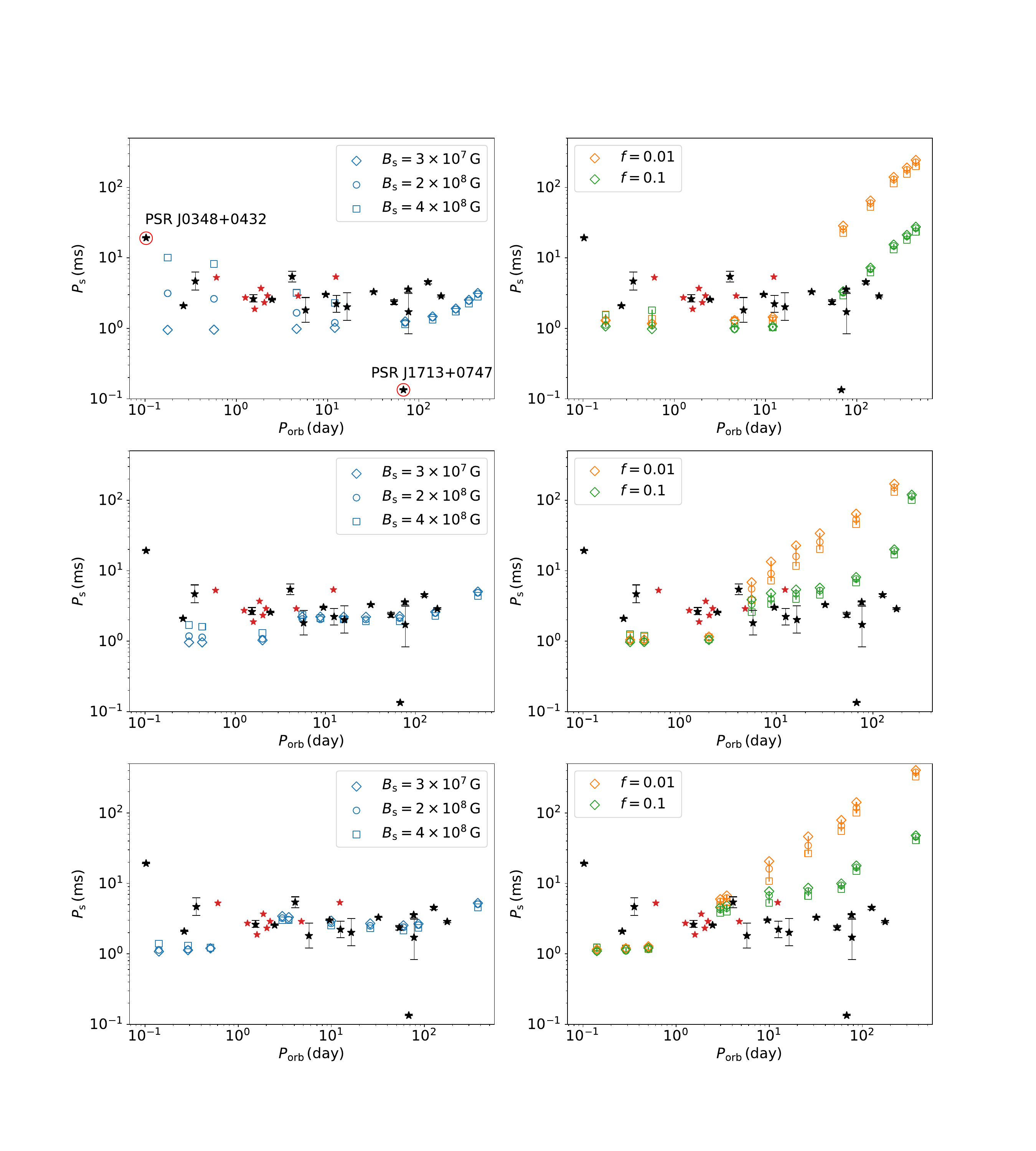}
		\caption{Comparison of the calculated spin periods with observations in the $M_{2}-P_{\rm s}$ and $P_{\rm orb}-P_{\rm s}$ diagrams. The initial values are $M_2=1.5\,\msun$ and $M_{\rm NS}=1.3\,\msun$. Panels from top to bottom show the results using MB1, MB2 and MB3 prescription, respectively. The corresponding initial orbital period $P_{\rm orb}$ are (0.55, 0.57, 0.7, 1.0, 5.0 ,10.0, 20.0, 30.0 and 40.0 day), (0.66, 0.7, 1.0, 5.0, 10.0, 15.0, 20.0, 30.0, 50.0, and 100.0 day), and (0.72, 0.8, 0.9, 7.0, 10.0, 20.0, 30.0, 40.0, 50.0 and 100.0 day), respectively. The diamond, circle and square points represent the results with $B_{\rm s}=3\times10^7, ,2\times10^8\, \rm{and}\,4\times10^8\,\rm G$, respectively and the blue (on the left column), yellow, and green (on the right column) colors the `stable' model, the `DI' models with $f=0.01$ and $f=0.1$, respectively. The red and black stars (with error bars) denote the observationally derived periods without and with considering the spin-down during radio MSP phase, respectively. }\label{fig:fig3}
	\end{figure*}

\begin{figure*}
		\includegraphics[width=1.0\linewidth]{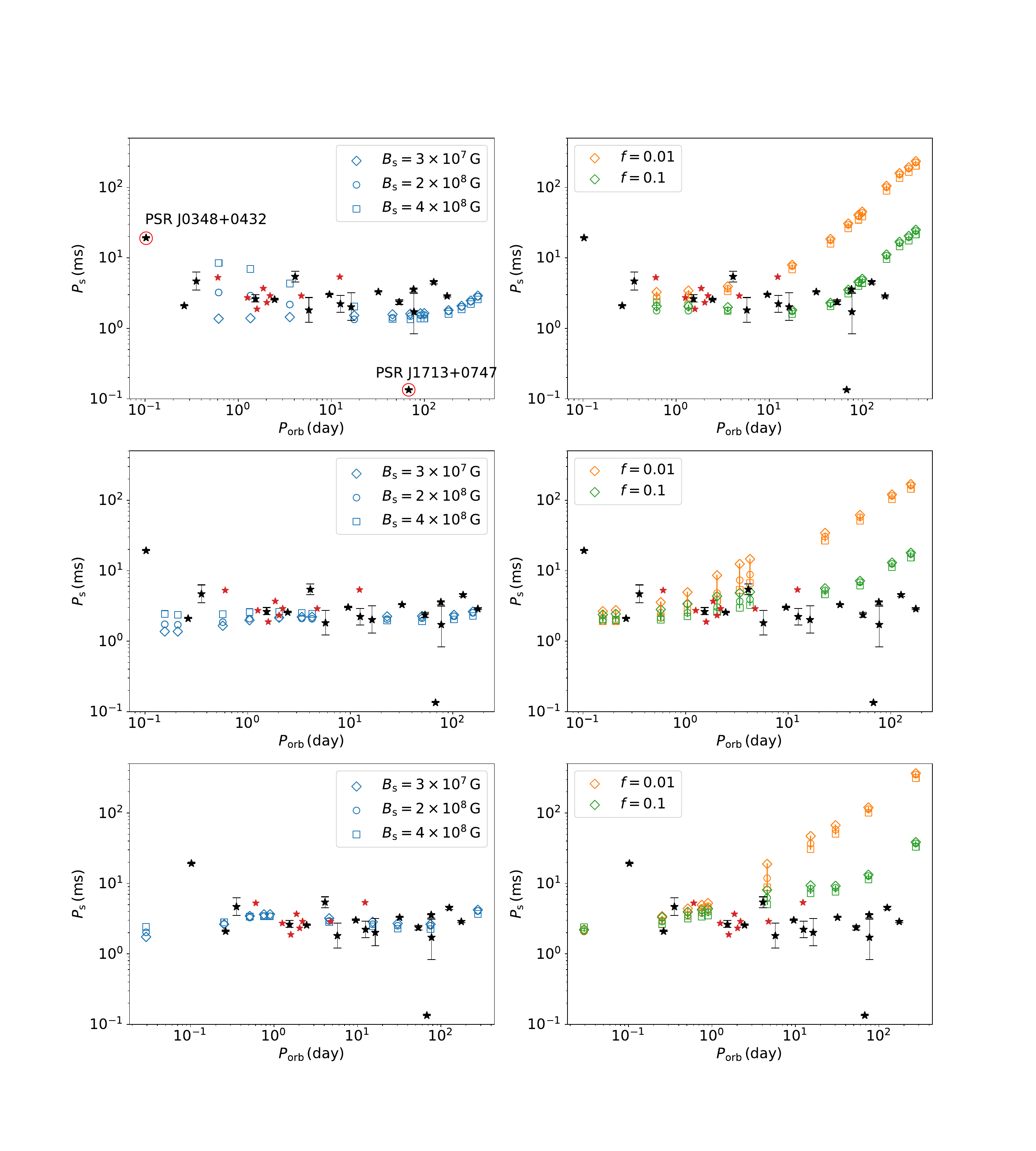}
		\caption{Same as Fig. \ref{fig:fig3}, but for initial donor star mass $M_2=1.0\,\msun$. For MB1, MB2, and MB3 prescriptions, the initial orbital periods are (2.86, 2.9, 3.0, 3.5, 5.0, 7.0, 9.0, 10.0, 20.0, 30.0, 40.0, 50.0 day), (3.34, 3.4, 4.0, 5.0, 7.0, 9.0, 10.0, 20.0, 30.0, 40.0, 50.0 day), and (4.4, 5.0, 7.0, 9.0, 10.0, 20.0, 30.0 40.0, 50.0, 100.0 day), respectively. }\label{fig:fig4}
	\end{figure*}

	We assume the final spin periods of the NSs in LMXBs to be the rebirth spin periods for MSPs, and plot them as a function of the orbital period in Figs.~\ref{fig:fig3} and \ref{fig:fig4}, in which $M_{2}=1.5\,\msun$ and $1.0\,\msun$, respectively. The NS magnetic fields are taken to be $3\times10^7-4\times 10^8\,\rm G$. In the upper, middle, and lower panels we adopt MB1, MB2, and MB3 prescriptions, respectively. The calculated results are compared with observationally derived rebirth spin periods for 25 MSP+He WD binaries in the Galactic plane, for which the cooling times of the WDs are available from optical observations\footnote{Most of the parameters are adopted from Australia Telescope National Facility (ATNF) \citep{Manchester2005AJ....129.1993M} catalogue and the references therein.}. Their parameters are listed in Table~\ref{tab:tab1}. We also list the four MSP+CO WDs with the cooling times measured for completeness in Table~\ref{tab:tab1}. Assume that the WD cooling time is equal to the spin-down time of the recycled pulsar, we can derive the rebirth spin periods of the pulsars based on the magnetic dipole radiation model. They are plotted with black stars. In some cases, taking the WD cooling ages as the NS spin-down ages is inapplicable for the standard magnetic dipole radiation, and we instead use their current periods as the rebirth periods. These sources are marked with red stars and marked in Table~\ref{tab:tab1} with a star symbol.
	For PSR J1713+0747, equating the estimated cooling age ($\simeq 9.1\,\rm Gyr$) of the WD companion with the pulsar's age yields an unreasonably small rebirth period (less than 1 ms). This implies that either standard magnetic dipole radiation does not apply for this pulsar \citep{Tauris2012MNRAS.425.1601T, Oliveira2018JCAP...11..025O} or the cooling age of the WD is significantly overestimated.
	
	Figs.~\ref{fig:fig3} and \ref{fig:fig4} show that, with increasing MB efficiency (MB1 $\rightarrow$ MB2 $\rightarrow$ MB3), a specific LMXB evolves to be a MSP+He WD binary with narrower orbits and less massive WDs. For instance, in the `stable' model, a binary with the initial parameters ($M_2$, $M_{\rm NS}$, $P_{\rm orb}$) = (1.0 $\msun$, 1.3 $\msun$, 10.0 $\rm day$) evolves to a MSP+He WD binary of (0.31 $\msun$, 1.50 $\msun$, 99.68 $\rm day$), (0.21 $\msun$, 1.44 $\msun$, 4.25 $\rm day$), and (0.18 $\msun$, 1.38 $\msun$, 0.90 $\rm day$) under the MB1, MB2, and MB3 prescriptions, respectively. The reason is that more efficient angular momentum loss enables the donor star to overflows its RL earlier, and hence leaves a less massive WD core. Besides, enhanced MB predicts faster spinning NSs in relatively compact binaries ($P_{\rm orb}< 1$ day) in the `stable' model.
	
	The spin periods differ remarkably in the `stable' (blue symbols on the left column) and `DI' (yellow and green symbols on the right column) models. In the `stable' model, the MSPs formed from short orbit period systems reach the equilibrium periods and remain basically unchanged during the long (a few Gyrs) mass transfer phase, and then experience additional spin-down during the decoupling phase; for wide systems, because of fast mass transfer, the magnetospheric radius is always close to the the NS radius, so the spin period evolution is not sensitively dependent on the magnetic field strength. The NSs keep spinning up toward the equilibrium spin period. In the `DI' model, for short orbit period systems, transient accretion results in shorter $P_{\rm s}$ because $\Rm$ during outbursts is considerably smaller than in the `stable' model; for wider systems, unstable mass transfer makes the NSs to accrete less mass, resulting in longer spin periods. That is why smaller duty cycle also leads to longer final spin periods. This helps reproduce the measured spin period distribution in wide MSP+He WD binaries, which seem to be longer than predicted by the `stable' model.
	
\begin{figure}
		\includegraphics[width=1.0\linewidth]{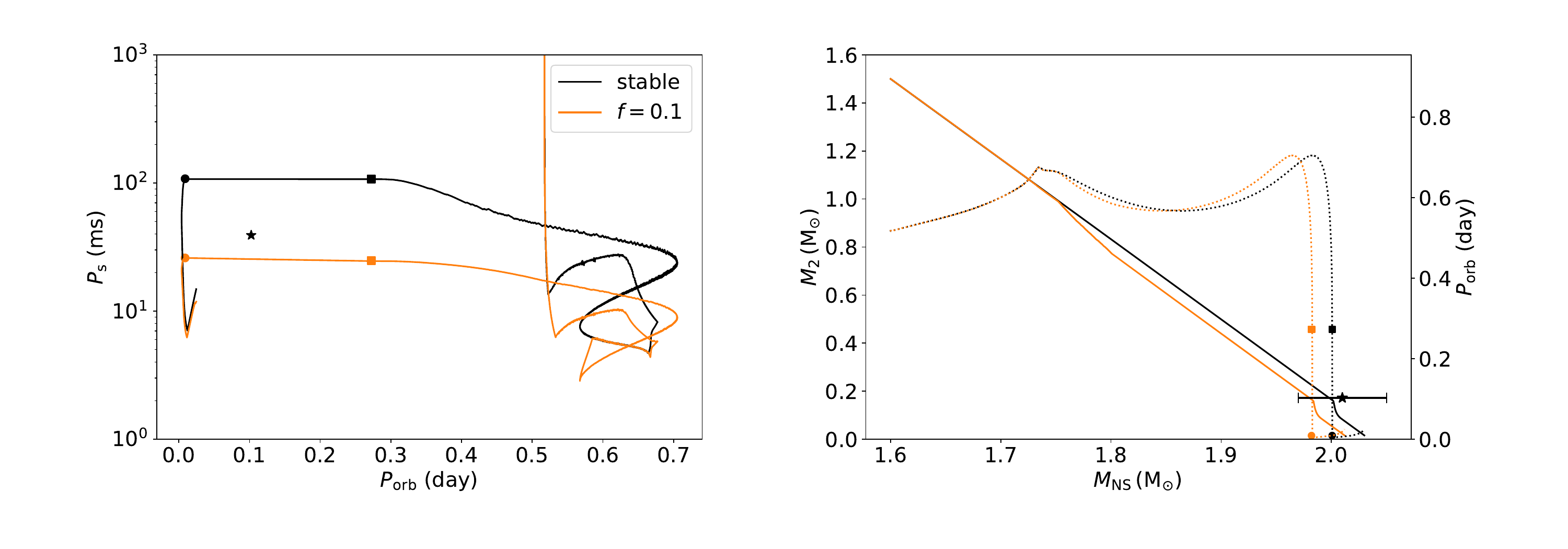}
		\caption{A possible model that can reproduce the observed parameters of PSR J0348+0432 binary (marked with $\star$). Left panel: Evolution of the spin period with orbital period. Right panel: Evolution of the donor star mass (solid lines) and orbital period (dotted lines) with NS mass. In each panel, the solid squares and circles in each track mark the termination of the first mass transfer phase and the onset of the second mass transfer phase, respectively. The initial parameters are ($M_2$, $M_{\rm NS}$, $P_{\rm orb}$, $B_{\rm s}$) = (1.5 $\msun$, 1.6 $\msun$, 0.52 day, $3\times10^9$ G). The black and yellow lines in each panel stand for the cases of the `stable' and `DI' ($f=0.1$) model, respectively. The MB1 prescription is used.}\label{fig:fig5}
	\end{figure}

	There is an outlier in Figs.~\ref{fig:fig3} and \ref{fig:fig4}, that is, PSR J0348+0432, which is a MSP with relatively long spin period of 39.123 ms. We notice that the MSP is a $2.01(\pm {0.04})\,\msun$ NS accompanied with a $0.172(\pm{0.003})\,\msun$ He WD in a very close ($P_{\rm orb}=0.1024$ day) orbit \citep{Antoniadis2013Sci...340..448A}.
	The massive NS, narrow orbit, and low-mass companion suggest that the NS may be born massive and have experienced efficient mass accretion and angular momentum loss during the LMXB evolution. To reproduce the parameters of PSR J0348+0432, we construct a model with the initial parameters ($M_{2},\,M_{\rm NS},\,P_{\rm orb},\,B_{\rm s}$) = ($1.5\,\msun,\,1.6\,\msun,\,0.52\,\rm{day},\,3\times10^9\,\rm{G}$), and demonstrate the calculated results in Fig.~\ref{fig:fig5}. The left and right panels show the evolution in the $P_{\rm s}-P_{\rm orb}$ plane and in the $M_2 (P_{\rm orb})-M_{\rm NS}$ plane, respectively. We present the results in both the `stable' and `DI' (with $f=0.1$) models. It is seen from the $P_{\rm s}-P_{\rm orb}$ panel that $P_{\rm s}$ lies between 25 ms and 108 ms (depending on the value of $f$) when $P_{\rm orb}$ decays to $0.1024$ day. In the $M_2 (P_{\rm orb})-M_{\rm NS}$ panel, for both the `stable' and `DI' models, the LMXB phase terminates (marked with the solid square) at the orbital period of $0.266$ day. Then the orbit decays to $0.1024$ day by gravitational wave radiation. To evolve to the observed state, it takes the proto-He WD about 2.7 Gyr to contract and settle on the cooling track, and another 2.3 Gyr to cool down. The current masses of the NS and the He WD are $\sim 1.96-2.0\,\msun$ and $\sim 0.164\,\msun$ respectively, in line with the observations. After about $0.5\,\rm Gyr$, the He WD will fill its RL once more, and the binary will evolve to be an UCXB.
	
	We note that the observationally inferred rebirth periods of MSPs are subject to the uncertainty in the WD cooling models, which mainly originates from the amount of residual nuclear burning. It is set by the thickness of the hydrogen envelope ($\sim 10^{-2}\,\msun$) above the He core before settling on the cooling track. For WDs massive than $\sim 0.17-0.20\,\msun$, the hydrogen layer burns unsteadily and leads to thermal flashes which can reduce the envelope mass (down to $\sim 10^{-3}\,\msun$), thus accelerate cooling, while for those below the threshold, steady hydrogen burning serves as the main energy source, prolonging the cooling time \citep{Webbink1975MNRAS.171..555W, Alberts1996Natur.380..676A, Driebe1998A&A...339..123D, Sarna1999ASPC..169..400S, Althaus2001MNRAS.323..471A}. The stationary hydrogen shell burning remains dominant for He WDs, which substantially prolongs their cooling times. And the He WD models with thick hydrogen envelopes would predict long cooling age for WD component. Hydrogen shell flashes caused RLOF mass loss as well as the pulsar irradiation of the proto He WD can influence the amount of residual hydrogen layer \citep{Iben1986ApJ...311..742I, Ergma2001MNRAS.321...71E}. Besides, the initial metallicity of the progenitors also affects the threshold mass for flashes - the lower metallicity, the lager threshold mass \citep{Antoniadis2012MNRAS.423.3316A, Serenelli2002MNRAS.337.1091S}.
	
	%Although the cooling age of a WD suffers from uncertainties, it is the only reliable age indicator of a pulsar binary system.
	%Increasing the number of MSP with optical counterparts identifications is important for understanding the nature and origin of these systems.
	
	\section{Conclusions}\label{sec:conclusion}
	
	Although the recycling scenario is widely accepted to be responsible for the formation of binary MSPs, the detailed processes are still not well understood. During the mass transfer via RLOF, accretion onto a magnetized NS is regulated by the coupling between the donor star and its RL, and by the properties of the accretion disc and its interaction with the NS. Thus, the traditional picture based on long-term, smooth accretion may not adequately reflect the real situation of the recycling process.
	
	In this work, we adopt three different MB prescriptions and consider transient accretion caused by the instability in the accretion disc to follow the spin evolution of NSs in LMXBs. We use the cooling age of the WDs in the observed MSP+He WD binaries as a proxy of the MSP's spin-down age, to derive the rebirth spin periods and make comparison with our theoretical predictions. Our conclusions are summarized as follows.
	
	1. We find that more efficient MB can produce faster MSPs in narrow MSP+He WD binaries in the `stable' model. However, if considering the effect of disc instability, there is not remarkable difference in the spin evolution in different MB models. Thus, the final spin periods do not sensitively depend on the MB efficiency.
	
	2. The accreting NSs all experience spin-down during the decoupling phase and deviate from the spin equilibrium. In the `stable' model, spin-down is more efficient in narrow systems than in wide systems, because of the longer duration of the decoupling process. In the `DI' model, the spin evolution shows a more complicated tendency than in previous studies. The overall feature is that transient accretion makes the NSs more difficult to reach the equilibrium spin. In narrow systems, the final spin periods of the accreting NSs are shorter than in the `stable' model because of more efficient spin-up during the outburst phases, but in wide systems, transient accretion results in slower spins than persistent accretion, because of more mass loss and smaller spin-up torque.
	
	\section*{Acknowledgments}\label{sec:acknowledge}
	This work was supported by the Natural Science Foundation of China under grant No. 12041301 and 12121003, and the National Key Research and Development Program of China (2021YFA0718500).
	We acknowledge use of the following PYTHON packages: ASTROPY \citep{Astropy2013A&A...558A..33A}, MATPLOTLIB \citep{Hunter2007CSE.....9...90H}, NUMPY \citep{van2011CSE....13b..22V} and SciPY \citep{Virtanen2020NatMe..17..261V}. %and CORNER \citep{Foreman2016JOSS....1...24F}.
	
	\section*{Data Availability}
	All data underlying this article will be shared on reasonable request to the corresponding authors.
	%%%%%%%%%%%%%%%%%%%%%%%%%%%%%%%%%%%%%%%%%%%

	\begin{sidewaystable*}[!hthp]
		\vspace{18.cm}
		\caption{Observed MSP+WDs in the Galactic plane with the WD's cooling time measured. The meanings of the parameters are as follows: $P_{\rm orb}$ and $P_{\rm s}$ - the current orbital and spin periods respectively, $\dot{P}_{\rm s}$ and $B_{\rm s}$ the ispin period derivative and the characteristic surface magnetic field strength after eliminating the effect of transverse motion \citep{Shklovskii1970SvA....13..562S}, $M_2$ - the WD masses. $\tau_{\rm c}$ - the characteristic age of MSPs,  $\tau_{\rm WD}$ - the WD cooling time. The corresponding references are listed in the last column.}\label{tab:tab1}
		\centering
		\small
		\begin{longtable}{ccccccccccc}
			\hline
			Pulsar& Type&  $P_{\rm orb}$& $M_{\rm NS}$&$M_{\rm 2}$&  $P_{\rm s}$& $\dot{P}_{\rm s}$&$B_{\rm s}$&$\tau_{\rm c}$&  $\tau_{\rm WD}$& Ref.\\
			& &  $(\rm day)$& $(\msun)$& $(\msun)$& (ms)&($\rm ss^{-1}$) &(G)&$(\rm {Gyr})$&  $(\rm {Gyr})$& \\\hline
			$\rm J0348+0432$& He&0.1024&$2.01\pm{0.04}$&$0.172\pm{0.003}$&	39.123& 2.35e-19&3.07e09&2.64&$\sim 2$&(1)\\\hline	
			
			$\rm J0751+1807$& He&	0.2631&$1.64\pm{0.15}$&	$0.16\pm{0.01}$&	3.479&  6.03e-21&1.47e08&9.14&$\sim 5.84$&(2)\\\hline
			
			$\rm J1738+0333$& He&	0.3548&	$1.47_{-0.06}^{+0.07}$&$0.181_{-0.005}^{+0.007}$&	5.85& 2.25e-20& 3.68e08&4.11&$\sim0.03-3$&(3)\\\hline
			
			$\rm J1012+5307^*$& He&0.6047&$1.72\pm{0.16}$&$0.165\pm{0.015}$&	5.256&  1.12e-20&2.46e08&7.41&$8\sim 9$&(4)\\\hline
			
			$\rm J0621+2514^{a,b,*}$& He&1.2564&1.35&$0.1715_{-0.0246}^{+0.1957}$&2.722&  2.48e-20& 2.63e08&1.74&$\lesssim 2$&(5)\\\hline
			
			$\rm J1909-3744$& He&1.5334 &$1.438\pm{0.024}$&$0.2038\pm{0.0022}$ &2.947&2.84e-21&9.26e07&16.4&$0.5\sim7$&(6)\\\hline
			
			$\rm J0034-0534^{b,*}$& He&	1.5893&	1.35&$0.1646_{-0.0235}^{+0.1869}$&	1.877&  4.06e-21&8.84e07&7.32&$7.0\pm{1.0}$& (7)\\\hline
			
			$\rm J1231-1411^{b,*}$& He&1.8601&1.35&	$0.2153_{-0.0314}^{+0.2549}$&	3.684&  8.19e-21&1.76e08&7.13&$\sim 5 - 8$&(8)\\\hline
			
			$\rm J0218+4232^{b,*}$& He&	2.0288&$1.6_{-0.7}^{+2.2}$&	$0.21_{-0.04}^{+0.17}$&2.323&  7.66e-20&4.27e08&4.80&$\sim 0.7$&(9)\\\hline
			
			$\rm J2017+0603^*$& He&2.1985&$2.4_{-1.4}^{+3.4}$&$0.32_{-0.16}^{+0.44}$&2.896& 7.94e-21& 1.53e08&5.78&$9\sim12$&(10)\\\hline
			
			$\rm J2317+1439$& He& 2.4593&$3.4_{-1.1}^{+1.4}$& $0.39_{-0.10}^{+0.13}$&3.445& 2.24e-21&8.89e07&24.4&$10.9_{-0.3}^{+0.3}$&(11)\\\hline
			
			$\rm J1045-4509^b$& He& 	4.0835&	1.35&	$0.1857_{-0.0267}^{+0.2147}$&	7.474& 1.73e-20&3.64e08&6.86&$\gtrsim2-5$&(12)\\\hline
			
			$\rm J0740+6620^*$& He& 	4.7669&$2.08\pm{0.07}$&$0.253_{-0.005}^{+0.006}$&	2.886&  3.65e-21&1.04e08&12.5&$\gtrsim 5$&(13)\\ \hline
			
			$\rm J0437-4715$& He&	5.7410&$1.44\pm{0.07}$&$0.224\pm{0.007}$&	5.757& 1.37e-20 &2.85e08&6.64&$6.0\pm {0.5}$&(14)\\\hline
			
			$\rm J1400-1431^b$& He&	9.5475&	1.35&$0.3102_{-0.0466}^{+0.3967}$&	3.084&  3.16e-22& 3.16e07&155&$\sim 5-9$&(15)\\\hline
			
			$\rm B1855+09^*$& He& 12.3272&$1.37_{-0.10}^{+0.13}$&$0.244_{-0.012}^{+0.014}$&	5.362& 1.73e-20&3.08e08&4.92&$3.757_{-1.514}^{+2.130}$&(16)\\\hline
			
			$\rm J1630+3734^b$& He&	12.5250&1.35&$0.2738_{-0.0407}^{+0.3401}$&	3.318& 8.30e-21& 1.68e08&6.34&$\sim 2-5$&(17)\\\hline
			
			$\rm J1741+1351$& He&16.3353&$1.87_{-0.69}^{+1.26}$&$0.32_{-0.09}^{+0.15}$&	3.747& 2.82e-20& 3.29e08&2.11&$\sim1-2$&(18)\\\hline
			
			$\rm J2234+0611$& He&	32.0014&$1.353_{-0.017}^{+0.014}$ &$0.298_{-0.012}^{+0.015}$&3.577& 5.82e-21& 1.46e08&9.74&$\sim1.5$&(19)\\\hline
			
			$\rm J0614-3329$& He & 53.5846&1.35& $0.24\pm{0.04}$&	3.149& 1.75e-20&2.38e08&2.85&$\sim1.0- 1.5$&(20)\\\hline
			
			$\rm J1713+0747$& He&	67.8251&$1.35\pm {0.07}$&	$0.292\pm{0.010}$&	4.570& 7.95e-21& 1.93e08&9.10&$\sim 9.1$&(21)\\\hline
			
			$\rm J2019+2425^b$& He&76.5116&1.35& $0.3643_{-0.0557}^{+0.4855}$&3.935& 1.32e-21&  7.30e07&47.2&$\lesssim8$&(22)\\\hline
			
			$\rm J2042+0246^b$& He&77.2006&1.35 &$0.2161_{-0.0315}^{+0.2561}$&	4.534& 1.10e-20& 2.26e08&6.52&$5.6_{-1.2}^{+0.9}$&(23)\\\hline
			
			$\rm J2302+4442$& He& 125.9353&$3.1_{-2.0}^{+2.7}$&$0.52_{-0.19}^{+0.25}$&	5.192& 1.35e-20& 2.68e08&6.09&$1\sim2$&(24)\\\hline
			
			$\rm J1640+2224^b$& He&	175.4607&1.35&	$0.2903_{-0.0433}^{+0.3656}$&	3.163& 1.27e-21& 6.42e07&39.4&$7\pm{2}$&(25)\\\hline
			
			$\rm B0655+64^b$& CO&	1.0287&1.35&$0.7966_{-0.1362}^{+1.4049}$&	195.671&  6.66e-19&1.15e10&4.66&$\sim 2$&(26)\\\hline
			
			$\rm J2145-0750$& CO& 6.8389&$1.8\pm{0.4}$& $0.9\pm{0.05}$&	16.052& 2.50e-20&6.41e08&10.2&$4.4\pm{0.2}$&(27)\\\hline
			
			$\rm J1022+1001$& CO&	7.8051&	$1.7\pm{0.3}$&$0.92\pm{0.05}$&	16.453& 3.64e-20& 7.83e08& 7.16& $2.7_{-0.2}^{+0.1}$&(28)\\\hline
			
			$\rm B0820+02^b$& CO&1232.4040&1.35&$0.2259_{-0.0330}^{+0.2700}$&	864.873& 1.04e-16&3.04e11&0.131&	$0.221\pm{0.011}$&(29)\\\hline		
			
		\end{longtable}%}
	\renewcommand{\thefootnote}{}\footnotetext{$^a$: Sources with $\dot{P}_{\rm s}$ and $B_{\rm s}$ without subtracting the kinematic effects because of the non detection of the distance or proper motion. Unless the transverse velocity is particularly large, the kinematic contribution to $\dot{P}_{\rm s}$ is negligible.}
	\renewcommand{\thefootnote}{}\footnotetext{$^b$: Sources where $M_{\rm NS}$ are derived from mass function, with orbital inclination $i=60^{\circ}$ and pulsar mass of 1.35 $\msun$, the lower/upper error bars corresponding to $i=90^{\rm o}$/$26^{\rm o}$, respectively.}
	\renewcommand{\thefootnote}{}\footnotetext{$*$: Sources inapplicable to magnetic dipole radiation theory if assuming $\tau_{\rm WD}$ as their spin-down time.}
	\renewcommand{\thefootnote}{}\footnotetext{References:
		(1)\cite{Antoniadis2013Sci...340..448A}($M_{\rm NS}$, $M_2$, $\tau_{\rm WD}$); (2)\cite{Desvignes2016MNRAS.458.3341D}($M_{\rm NS}$, $M_2$);\cite{Panei2007MNRAS.382..779P}($\tau_{\rm WD}$); (3)\cite{Antoniadis2012MNRAS.423.3316A}($M_{\rm NS}$, $M_2$); \cite{Kilic2018MNRAS.479.1267K}($\tau_{\rm WD}$); (4)\cite{Mata2020MNRAS.494.4031M}($M_{\rm NS}$,$M_2$); \cite{Nelson2004ApJ...616.1124N}($\tau_{\rm WD}$); (5) \cite{Bobakov2023MNRAS.524.3357B}($\tau_{\rm WD}$); (6)\cite{Jacoby2005ApJ...629L.113J}($M_{\rm NS}$, $M_2$); \cite{Liu2020MNRAS.499.2276L}($\tau_{\rm WD}$); (7)\cite{Althaus2001MNRAS.323..471A}($\tau_{\rm WD}$); (8)\cite{Bailes2003ApJ...595L..49B}($M_{\rm NS}$, $M_2$),\cite{Bassa2016MNRAS.455.3806B}($\tau_{\rm WD}$); (9)\cite{Bassa2003A&A...403.1067B}($M_{\rm NS}$, $M_2$, $\tau_{\rm WD}$); (10)\cite{Fonseca2016ApJ...832..167F}($M_{\rm NS}$, $M_2$); \cite{Bassa2016MNRAS.455.3806B}($\tau_{\rm WD}$); (11)\cite{Dai2017ApJ...842..105D}($M_{\rm NS}$, $M_2$,$\tau_{\rm WD}$); (12)\cite{Bobakov2019JPhCS1400b2023B}($\tau_{\rm WD}$); (13)\cite{Fonseca2021ApJ...915L..12F}($M_{\rm NS}$, $M_2$);\cite{Beronya2019MNRAS.485.3715B}($\tau_{\rm WD}$); (14)\cite{Reardon2016MNRAS.455.1751R}($M_{\rm NS}$,$M_2$); \cite{Durant2012ApJ...746....6D}($\tau_{\rm WD}$); (15)\cite{Swiggum2017ApJ...847...25S}($\tau_{\rm WD}$); (16)\cite{Arzoumanian2018ApJS..235...37A}($M_{\rm NS}$, $M_2$); \cite{Panei2007MNRAS.382..779P}($\tau_{\rm WD}$); (17)\cite{Kirichenko2020MNRAS.492.3032K}($\tau_{\rm WD}$); (18)\cite{Fonseca2016ApJ...832..167F}($M_{\rm NS}$, $M_2$); \cite{Kirichenko2020MNRAS.492.3032K}($\tau_{\rm WD}$); (19)\cite{Stovall2019ApJ...870...74S}($M_{\rm NS}$, $M_2$);\cite{Antoniadis2016ApJ...830...36A}($\tau_{\rm WD}$);  (20)\cite{Bassa2016MNRAS.455.3806B}($M_2$,$\tau_{\rm WD}$); (21)\cite{Arzoumanian2018ApJS..235...37A}($M_{\rm NS}$, $M_2$); \cite{Althaus2001MNRAS.323..471A}($\tau_{\rm WD}$); (22)\cite{Lundgren1996ASPC..105..497L}($\tau_{\rm WD}$); (23)\cite{Kirichenko2020MNRAS.492.3032K}($\tau_{\rm WD}$); (24)\cite{Kirichenko2018MNRAS.480.1950K}($M_{\rm NS}$, $M_2$, $\tau_{\rm WD}$); (25)\cite{Lundgren1996ASPC..105..497L}($\tau_{\rm WD}$); (26)\cite{van-Kerkwijk2005ASPC..328..357V}($\tau_{\rm WD}$); (27)\cite{Deller2016ApJ...828....8D}($M_{\rm NS}$, $M_2$, $\tau_{\rm WD}$); (28)\cite{Deller2016ApJ...828....8D}($M_{\rm NS}$, $M_2$,$\tau_{\rm WD}$); (29)\cite{Koester2000A&A...364L..66K}($\tau_{\rm WD}$).
	}
\end{sidewaystable*}
%%%%%%%%%%%%%%%%%% APPENDIX %%%%%%%%%%%%%%%%%%%%%%%%%%%%%%%%%%%%%%%

%%%%%%%%%%%%%%%%%%%%%%%%%%%%%%%%%%%%%%%%%%%%%%%%%%%%%%%%%%%%%

%%%%%%%%%%%%%%%%%%

\label{lastpage}

\end{document}